\documentclass[%
reprint,
onecolumn,
amsmath,amssymb,
aps,
prb,
]{revtex4-2}

\usepackage{graphicx}% Include figure files
\usepackage{dcolumn}% Align table columns on decimal point
\usepackage{bm}% bold math
\usepackage[T2A]{fontenc}
\usepackage[utf8]{inputenc}

\usepackage[usenames]{color}
\usepackage{colortbl}
\usepackage{amsmath}
\usepackage{amssymb}
\usepackage{physics}

\begin{document}

\preprint{APS/123-QED}

\title{Disordered quantum antiferromagnetism in doped semiconductors: Density of states approach}% Force line breaks with \\

\author{N. A. Bogoslovskiy}
\author{P. V. Petrov}
 \email{pavel.petrov@gmail.com}
\author{N. S. Averkiev}
\affiliation{Ioffe Institute, Russian Academy of Science, 194021 St. Petersburg, Russia}

\date{\today}% It is always \today, today,

\begin{abstract}
We present a theoretical study of the exchange interaction in a system of spatially disordered magnetic moments.
A typical example of such a system is an impurity semiconductor, whose magnetic properties
are associated with the exchange interaction of the impurity atoms. In this study, we consider the
case of antiferromagnetic exchange interaction, which we describe by the Heisenberg Hamiltonian. To
calculate the magnetic properties of the disordered system, we employ the joint density of states
method. Our calculations demonstrate a good quantitative agreement with the experimental dependence of the Si:P
magnetic susceptibility on temperature in a wide range of impurity concentrations and temperatures.
The obtained results may be important for a deeper understanding of the metal-dielectric transition nature.
%\begin{description}
%\end{description}
\end{abstract}

\maketitle
\section{Introduction}

The study of disordered quantum systems has long been a topic of interest for researchers due to the
inherent complexity and scientific significance of these systems. These studies are currently being
actively conducted both experimentally~\cite{armstrong2025experimental, PhysRevB.110.134429, PhysRevResearch.6.L032010}
and theoretically~\cite{PhysRevB.111.104202, PhysRevLett.134.086501, PhysRevB.110.024409, PhysRevB.107.134418, PhysRevB.103.125120}.
An interplay of disorder and antiferromagnetism is studied in various systems such as spin chains, ladders and lattices.
Most of this studies focus on periodic structures, while systems with structural disorder are much less investigated.

Impurities in doped semiconductors represent a prominent example of such a system. 
The antiferromagnetic exchange interaction of electrons localized at impurity centers affects the spin ordering, which in turn gives rise to a deviation of their magnetic susceptibility from the Curie law.
This effect is particularly pronounced in the vicinity of the metal-insulator transition, where the mean distance between impurity centers is merely several times greater than the radius of the electron wave function on the impurity~\cite{andp.201100034, Mobius02012019, KETTEMANN2023169306}.
Usually the properties of such systems at low temperatures are theoretically described using the strong disorder renormalization group approach~\cite{IGLOI2005277, PhysRevB.94.174442, PhysRevB.110.054204}.
This approach was initially proposed for the study of quantum chains~\cite{ma1979random} and was subsequently successfully applied by Bhatt and Lee to the investigation of magnetic properties of doped semiconductors~\cite{PhysRevLett.48.344}.
At low temperatures this model predicts a power dependence of the magnetic susceptibility on temperature
$\chi \sim T^{-\alpha}$, where $\alpha$ is a phenomenological parameter dependent on the impurity concentration.

The model demonstrates a good agreement with experimental results at temperatures below the characteristic energy of the exchange interaction between electrons localized on impurities \cite{PhysRevB.37.5531, Schlager_1997}.
At higher temperatures the measurements give the following dependence of the magnetic susceptibility on temperature $\chi \sim (T+T_c)^{-1}$ \cite{PhysRevB.24.244, PhysRevB.37.5531}.
To the best of our knowledge, a comprehensive and detailed quantitative description of the behaviour of magnetic susceptibility within the specified temperature range has yet to be provided.
The objective of this study is to propose a quantitative model that can describe the magnetic properties
of disordered antiferromagnetism in doped semiconductors using the density of states approach.
We used a similar approach in our previous work to study ferromagnetism and superparamagnetism in semiconductors~\cite{PhysRevB.109.024436}.

In this approach, the joint density of states, denoted by $g(E, M)$, is calculated as a function of the total energy of the spin system $E$ and the total magnetic moment $M$~\cite{PhysRevLett.96.120201}. 
The known joint density of states allows one to calculate the statistical sum, 
$Z(T, B) = \iint g(E, M) \exp\left(-\frac{E-BM}{T}\right)\,dE\,dM$, 
which is a function of the magnetic field, $B$, and the temperature, $T$. 
Moreover, within the framework of the standard thermodynamic approach, other parameters of the system can also be calculated.
In a pioneering work on the nature of ferromagnetism~\cite{Heisenberg1928}, Heisenberg initially proposed the idea of using the density of states to describe the magnetic properties of a spin system.
Attention was again drawn to this approach when it was shown that the density of states of spin
systems can be calculated with a reasonable accuracy through numerical simulations,
initially as a function of energy~\cite{PhysRevLett.86.2050}, and later as a joint density of
states, i.e. a function of energy and magnetic moment~\cite{PhysRevLett.96.120201,
Egorov_Kryzhanovsky_2024}.
Furthermore, the density of states can be found using the central limit theorem.
The total energy of the system is represented as the sum of the energies of individual spins, which are treated as independent random variables.
In this case, the total energy is distributed normally; and the mean energy and variance are functions of the magnetic moment.
It has been demonstrated that this method is applicable to the Ising model on a periodic lattice, although good agreement with the exact result is only possible in high-dimensional systems~\cite{e23121665}.
However, for a disordered system, the analytical approach based on the normal distribution demonstrates a good agreement with the results of numerical simulations in three-dimensional space~\cite{PhysRevB.109.024436, 10.18721/JPM.161.301}.
A deeper and rigorous discussion about applicability of the normal distribution in quantum statistics can be found in Khinchin's book
``Mathematical foundations of quantum statistics''~\cite{khinchin2013mathematical}.

In this paper, we apply the density of states approach to study a system of randomly distributed spins in three-dimensional space.
We consider the case of antiferromagnetic interaction of spins, which is described by the Heisenberg Hamiltonian.
First, we will describe how to apply the density of states method to doped semiconductors.
Next, the dependences of the average energy and the variance on the magnetic moment will be calculated.
The results of analytical calculations will be compared with the results of numerical calculations for systems containing from 2 to 16 spins randomly distributed in 3D space.
Then the magnetic susceptibility in relatively weak magnetic fields will be calculated. 
In order to demonstrate the validity of the density of states approach we aplly this approach the the Heisenberg $N$-spin chain, a well studied problem with well known results ~\cite{Bethe, PhysRev.135.A640}.
The calculated susceptibility will be compared with the experimentally measured temperature dependences of the magnetic susceptibility of phosphorus-doped silicon.
The exchange interaction of impurity atoms in silicon has recently attracted renewed attention due to its potential applications in quantum computing \cite{PhysRevB.72.085202, RevModPhys.85.961, PhysRevB.105.155158, PhysRevApplied.21.014038}.
Finally, the calculations of magnetic moment in strong magnetic fields will be presented and a comparison with experimental data will be made.

\section{Density of states approach for Heisenberg Hamiltonian}

We consider a system of $N$ randomly distributed atoms with spin $\frac{1}{2}$. Such a model can describe randomly distributed impurities
in semiconductor material.
The state of a spin system can be specified by the total magnetic moment and the projection of the magnetic moment onto a specified axis.
The energy of the system in a magnetic field is determined by the projection of the magnetic moment onto the direction of the magnetic field.
For the sake of brevity, the direction of the magnetic field will be designated as the $oz$ axis. 
The spins with projection $+ \frac{1}{2}$ onto the $oz$ axis will be designated 
as "spin-up" or $| \uparrow \rangle$, and spins with projection $- \frac{1}{2}$ 
will be designated as "spin-down" or $| \downarrow \rangle$.

For a large system of magnetic atoms, the macroscopic parameters of the system, such as magnetic susceptibility can be calculated using the statistical mechanics methods.
The state of a system is characterised by macroscopic parameters, and the probability of each macrostate depends on the number of associated microstates.

If the concentration of magnetic atoms is sufficiently high, each spin will interact with a considerable number of neighbors. In this case, the exchange energy of a single spin has a distribution that is approximately normal~\cite{bogoslovskiy2021spin, 10.18721/JPM.161.301}.
The total exchange energy is the sum of the single-spin energies. Consequently, the total exchange energy of a system of  $N \gg 1$ randomly distributed spins also has a normal distribution.
To find $g(E,M)$ for a system of $N$ spins, it is sufficient to find the average value and the variance of the exchange energy as a function of $M$.

In this paper we use the Heisenberg Hamiltonian to describe the exchange interaction between two spins
\begin{equation}
\vu{H} = \frac{1}{2} \sum_{i,j} J_{ij} \vu{S}_i \vu{S}_j - g \mu B \sum_i \vu{s}_{zi}
\label{H}
\end{equation}
Here $\mu$ is the Bohr magneton and $g$ is the spin g-factor.
In the first term $J_{ij}$ depends on the coordinates of the spin. In the second term $\vu{s}_z$ is the projection of the spin onto the $oz$ axis.
In this paper the dependence of $J_{ij}$ on the distance between spins $r_{ij}$ will be described by a hydrogen-like model~\cite{1964Gorkov, PhysRev.134.A362}

\begin{equation}
J(r) = J_0 \left( \frac{r}{a} \right)^{5/2} \exp \left( -\frac{2r}{a} \right)
\label{J_r}
\end{equation}
Here $a$ is the Bohr radius. Such a dependence will allow us to obtain analytical results.
Below we will discuss the applicability of equation~(\ref{J_r}) for the description of the exchange interaction between phosphorus impurities in silicon.

%Since we are considering systems with a large number of spins, it can be reasonably assumed that the total exchange energy will be close to the ensemble average.
%In order to calculate the average exchange energy, it is necessary to perform two separate averages: one over the coordinates and one over the projections of the spins onto the $oz$ axis.

In order to find the average exchange energy of the system, it is necessary to average over
coordinates and over the projections of the spins onto the $oz$ axis. In this paper we consider a
system of impurities in a semiconductor. The impurities are randomly distributed in the material
during growth, after that the spatial coordinates of the impurities do not change, however the spin
projections change with time.
Therefore, averaging over coordinates should be understood as ensemble averaging. And averaging over the directions of spins could be performed independently.
The averaging of the spin projections is performed as follows.
A system of two spins in a zero magnetic field has four states. 
The state $\mid \uparrow \uparrow \rangle$ with energy $+\frac{1}{4}$, the state $\mid \downarrow \downarrow \rangle$ with energy $+ \frac{1}{4}$, and two states with zero projection of the magnetic moment onto the $oz$ axis.
The symmetric state $\frac{\mid \uparrow \downarrow \rangle + \mid \downarrow \uparrow \rangle}{\sqrt 2}$ with energy $+ \frac{1}{4}$ and the antisymmetric state $\frac{\mid \uparrow \downarrow \rangle - \mid \downarrow \uparrow \rangle}{\sqrt 2}$ with energy $-\frac{3}{4}$.

The number of up spins can be expressed as $q = \frac{N}{2} + \frac{M}{g \mu}$. Among all $\frac{1}{2} N(N-1)$ pairs of spins, $\frac{1}{2} q(q-1)$ pairs have a projection of the magnetic moment equal to $1$ and an energy equal to $+\frac{1}{4}$.
$\frac{1}{2} (N-q)(N-q-1)$ pairs have a projection of the magnetic moment equal to $-1$ and an energy equal to $+\frac{1}{4}$.
The remaining $(N-q) q$ pairs of spins have a projection of the magnetic moment equal to $0$ and are equally distributed between states with symmetric and antisymmetric spin part of the wave function.
The average energy for such states is equal to $-\frac{1}{4}$.
Averaging over spins gives
\begin{equation}
\begin{aligned}
\label{sij}
\overline {\vu{S}_i \vu{S}_j} = & \frac{1}{4} \frac{q(q-1) + (N-q)(N-q-1) - 2q(N-q)}{N(N-1)} \\
=& \frac{1}{4}  \left(  m^2+ O\left( \frac{1}{N} \right) \right)
\end{aligned}
\end{equation}
Here $m$ is the dimensionless magnetic moment per spin $m = \frac{2 M}{g \mu N}$, which varies from $-1$ to $1$.
Next we consider the average energy of the exchange interaction between a single spin and all others $ \sum_{j} J_{ij} $ to average the exchange energy over the spin coordinates.
Given the random distribution of spins, we can transition from the sum over discretely located spins to the integral over the continuous magnetic moment with a density $n$.

\begin{equation} \label{j1}
\overline {J_1} = \int \limits_0^\infty {n J(r) 4\pi r^2 dr} =
\frac{945 \pi}{2^{8}} \sqrt {\frac {\pi}{2}} J_0 n a^3
\end{equation}

The mean value of the total exchange energy is
\begin{equation}
\overline {E} = \frac{N}{2} \overline {J_1} \overline {\vu{S}_i \vu{S}_j} =
\frac{1}{8} N \overline{J_1} \left(  m^2 + O\left( \frac{1}{N} \right) \right)
\label{mean_E}
\end{equation}

Next we will calculate the variance of the exchange energy.
\begin{equation} 
\nonumber
\sigma^2 = \overline {E^2} - {\overline {E}}^2 
\end{equation}

We substitute an explicit expression for the exchange energy
\begin{equation} 
\nonumber
\sigma^2 = \frac{1}{4} \overline {\sum \limits_{\substack{i, j \\ k, m}} J_{ij} J_{k l} \vu{S}_i \vu{S}_j \vu{S}_k \vu{S}_l } - 
\frac{1}{2} \overline {\sum \limits_{i, j} J_{i j}  \vu{S}_i \vu{S}_j} \frac{1}{2} \overline {\sum \limits_{k, l} J_{kl} \vu{S}_k \vu{S}_l}
\end{equation}

As stated above, here we consider a system of randomly oriented spins. This implies that there is no
correlation between the value of the exchange energy $ J_{ij}$ and the direction of the spin
$\vu{S}_i$.
Consequently, the averaging over coordinates and over spin directions should be carried out separately.

\begin{equation}
\begin{aligned} 
\sigma^2 = \frac{1}{4} \overline {\sum \limits_{\substack{i, j \\ k, l}} J_{ij} J_{kl}} \overline { \vu{S}_i \vu{S}_j \vu{S}_k \vu{S}_l } -  \\
\frac{1}{2} \overline {\sum \limits_{i, j} J_{ij}}  \overline {\vu{S}_i \vu{S}_j }  \frac{1}{2} \overline {\sum \limits_{k, l} J_{kl}} \overline {\vu{S}_k \vu{S}_l}
\label{sigma2}
\end{aligned}
\end{equation}

We will separately consider the terms for which all 4 indices $(i, j, k, l)$ are different, two
indices coincide and two pairs of indices coincide. First we consieder the case when all indices are
different.
The averaging of the spin component is conducted similar to (\ref{sij})
\begin{equation} 
\nonumber
\overline { \vu{S}_i \vu{S}_j \vu{S}_k \vu{S}_l} =\frac{1}{16} \left( m^4 - \frac {6m^2}{N} + \frac {6m^4}{N} + O\left( \frac{1}{N^2} \right) \right)
\end{equation}
The sum $\sum \limits_{i, j, k, l} J_{ij} J_{kl} $ can be considered as the product of two sums, $i, j$ and $k, l$. Since all indices are different,
the two sums may be averaged independently $\overline{J_{ij}J_{kl}} = \overline{J_{ij}}~\overline{J_{kl}}$.

\begin{equation} 
\nonumber
\frac{1}{2} \overline {\sum \limits_{i, j} J_{ij}} = \frac{1}{2}N\overline{J_1}
\end{equation}

The indices $k, l$ must not coincide with the indices $i,j$ from the first sum. Consequently, for each of the indices $k, l$ there are only $N-2$ potential values
\begin{equation} 
\nonumber
\frac{1}{2} \overline {\sum \limits_{k, l} J_{kl}} = \frac{1}{2}\frac{(N - 2)^2}{N}\overline{J_1} = \frac{1}{2} \left( N - 4 \right) \overline{J_1} + O\left( \frac{1}{N} \right)
\end{equation}

\begin{equation} 
\nonumber
\frac{1}{4} \overline {\sum \limits_{\substack{i, j \\ k, l}} J_{ij} J_{kl}} = \frac{1}{4} \left( N^2- 4N \right) \overline{J_1}^2 + O\left( 1 \right)
\end{equation}

We substitute these expressions into the variance (\ref{sigma2}). The terms of order $N^2$ cancel
and finally the terms with 4 different indices in (\ref{sigma2}) give
\begin{equation}
\frac {1}{16} N \overline{J_1}^2 (m^4-m^2) + O\left( 1 \right)
\label{4index}
\end{equation}

Next we consider the terms in (\ref{sigma2}) with exactly two matching indices. We will denote the
matching indices by $i$, and different indices by $j$ and $k$.
In the original notation, there are four possible options for matching indices: $(i = k, i = l, j = k, j = l)$.  Consequently, after redesignation, the multipliers $\frac{1}{4}$ and $\frac{1}{2}$ in (\ref{sigma2})
will be canceled.
In the second term of the expression, we rewrite the two sums over $i, j$ and $i, k$ as a total sum over three indices.
\begin{equation}
\nonumber
\overline {\sum \limits_{\substack{i, j, k \\ k \neq j}} J_{ij} J_{ik}} \overline {\vu{S}_i \vu{S}_j \vu{S}_i \vu{S}_k } - 
\overline {\sum \limits_{\substack{i, j, k \\ k \neq j}} J_{ij} J_{ik}} \overline {\vu{S}_i \vu{S}_j }  \;  \overline { \vu{S}_i \vu{S}_k }
\end{equation}

Averaging over the coordinates gives
\begin{equation}
\nonumber
\overline {\sum \limits_{\substack{i, j, k \\ k \neq j}} J_{ij} J_{ik}} = N \overline{J_1}^2 + O\left( 1 \right)
\end{equation}

In order to calculate the mean value of the spin part, it is necessary to determine the number of spin pairs with different projections of the magnetic moment depending on $m$, as in the case with four different indices. Up to terms of higher order in $N$ we obtain
\begin{equation}
\nonumber
\overline {\vu{S}_i \vu{S}_j \vu{S}_i \vu{S}_k } = \frac{1}{16} m^2 + O\left(  \frac{1}{N} \right)
\end{equation}

Finally
\begin{equation}
\begin{aligned}
\label{ijk}
\overline {\sum \limits_{\substack{i, j, k \\ k \neq j}} J_{ij} J_{ik}} \left(
\overline {\vu{S}_i \vu{S}_j \vu{S}_i \vu{S}_k } - \overline {\vu{S}_i \vu{S}_j } \;  \overline { \vu{S}_i \vu{S}_k } \right) \\
= \frac{1}{16} N \overline{J_1}^2 \left( m^2- m^4 \right) + O\left( 1 \right)
\end{aligned}
\end{equation}

Next, we consider the terms (\ref{sigma2}) with two pairs of coinciding indices.
A pair of spins can be in four quantum states. 
Similar to (\ref{sij}) we can express the probability of each state depending on the projection of the magnetic moment $m$ and obtain for the mean value for the spin component
\begin{equation} \label{2index}
\overline { \left( \vu{S}_i \vu{S}_j \right) ^2 }  -  \left( \overline {\vu{S}_i \vu{S}_j } \right) ^2 = 
\frac{1}{16} \left( 3 - 2 m^2 - m^4 \right)
\end{equation}

To average over coordinates, we first calculate the square of the exchange energy for two spins, labelled $i$ and $j$, provided that the distance between them does not exceed $R$
\begin{equation} 
\nonumber
\overline {J^2_{ij}(R)} = \frac{3}{4 \pi R^3} \int \limits_0^R J^2(r) 4 \pi r^2 dr
\end{equation}

We substitute the explicit expression for $J(r)$ and tend the upper limit to infinity since $J(r)$ exponentially decreases with $r$.
\begin{equation} 
\nonumber
\overline {J^2_{ij}} = \frac{3 J_0^2 a^3}{R^3 2^{16}} \int \limits_0^{\infty} \left(\frac{4r}{a} \right)^{7} \exp \left( - \frac{4 r}{a} \right) \,d {\left(\frac{4 r}{a} \right)}
\end{equation}

After integration we obtain
\begin{equation} 
\nonumber
\overline {J^2_{ij}} = \frac{3 \cdot 7! J_0^2 a^3}{2^{16} R^3}
\end{equation}

In our model, the arrangement of spins is independent, therefore the sum of the squares of the exchange energies between spin $i$ and all spins from a sphere of radius $R$ is equal to
\begin{equation} \label{J_square}
\overline {\sum \limits_{j} J^2_{ij}} = n \frac{4}{3}\pi R^3 \overline{J^2_{ij}} = \frac{7! \pi J_0^2 n a^3}{2^{14}}
\end{equation}

We sum (\ref{4index}-\ref{J_square}) and obtain the variance
\begin{equation}
\begin{aligned}
\sigma^2 = 
\frac{1}{2} \overline {\sum \limits_{i, j} J_{ij}^2} \frac{1}{16} \left( 3 - 2m^2 - m^4 \right) = \\
N \sigma_1^2 \left(1 - \frac{2}{3} m^2 - \frac{1}{3} m^4 \right) 
\label{variance_E}
\end{aligned}
\end{equation}

Here we introduce the notation $\sigma_1^2$, which is equal to the variance of the exchange energy at $m = 0$. 

\begin{equation} 
\sigma_1^2 = \frac{3}{32 N} \overline {\sum \limits_{i,j} J^2_{ij}} = \frac{3 \cdot 7!}{2^{19}} \pi J_0^2 n a^3
\label{variance_1}
\end{equation}

\section{Numerical simulation}
In order to verify the formulas (\ref{mean_E}) and (\ref{variance_E}) a numerical simulation was
conducted for a system of randomly distributed spins with Heisenberg interaction~(\ref{H}).
The spins were randomly placed within a cube with periodic boundary conditions.
Subsequently, the interaction matrix $J_{ij}$ was calculated for the hydrogen-like exchange interaction~(\ref{J_r}).

The spectrum of the system was calculated via exact numerical diagonalization of the Hamiltonian.
To accelerate the computations, we  developed a program in the Julia language~\cite{besard2018juliagpu}, 
which uses CUDA technology for parallel computing on the GPU.
We were able to compute the densities of states for a system of up to 16 spins.

%The calculation of the spectrum of a 16-spin system requires the diagonalization of the $2^{16}\times2^{16}$ matrix 
%and averaging of realizations of geometric disorder, which is a time-consuming process.
%Therefore, prior to the calculations, the first four moments of the distribution of single-spin energies $J_i = \sum_{i\neq j} J_{ij}$ averaged over $10^6$ realizations of disorder were calculated.
%Further, instead of averaging over a large number of implementations, we studied only those systems for which the distribution moments $J_i$ differed from the average by no more than a certain value $\delta J$, expressed in percentage. That is, rather than averaging the calculated densities of states $g(E, M)$ over coordinates, we calculated the densities of states for spin configurations that are close to the average ones.

A system of 16 spins is rather small, therefore the energy fluctuations in such a system are large.
To reduce the fluctuations, one can average the results over a large number of different
realisations. However, this requires considerable computational time. Therefore, in the present work
we calculated the first four moments of the single-spin energy distribution $J_i = \sum_{i\neq j} J_{ij}$ for $10^6$
different spin configurations. Then, instead of averaging over a large number of realisations, we
studied only those systems for which the distribution moments $J_i$ differed from the average by no
more than a certain value $\delta J$, expressed in percentage. That is, instead of averaging the
calculated densities of states $g(E, M)$ over the coordinates, we calculated the density of states for
one spin configuration, which is close to an average one.

\begin{figure*}[!htbp]
\centering
\includegraphics[width=\textwidth]{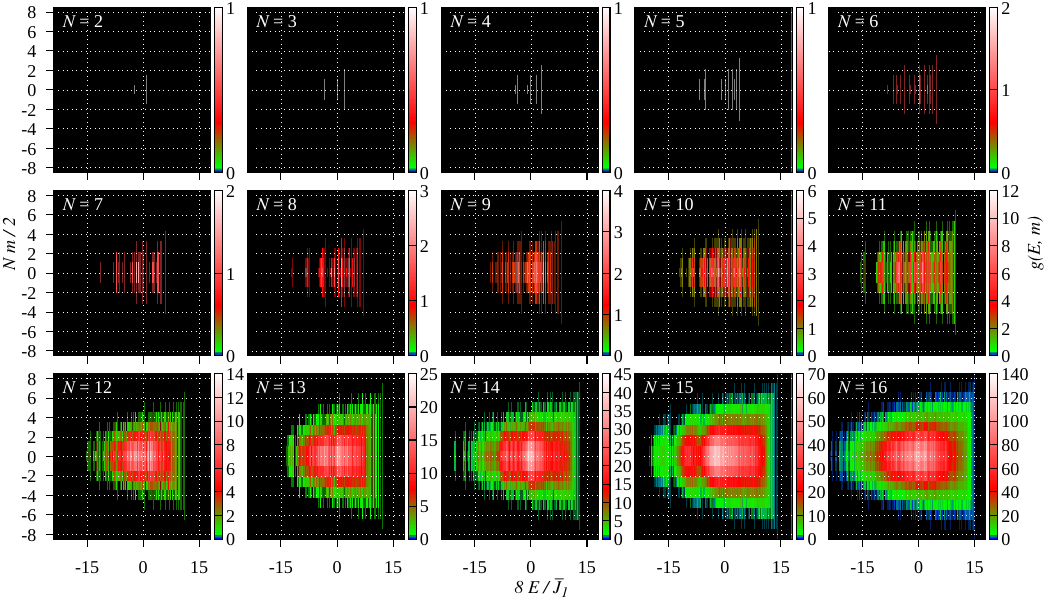}
\caption{Density of states $g(E, m)$ for a system of $N$ randomly distributed spins. The system size is shown in the figures.
$\delta J=1\%, na^3=0.01$}
\label{g_Em}
\end{figure*}

The figure~\ref{g_Em} shows the densities of states for different $N$. 
As expected, for $N=2$, $g(E, m)$ consists of two states: singlet with $m=0$ and triplet $m=0, \pm 1$.
As the size of the system increases, the form of $g(E, m)$ becomes more complicated. It has a strong maximum at
zero energy and a quite asymmetric dependence on energy. While temperature is high the states in the maximum of $g(E, m)$ are occupied.
At low temperature the system occupies states at the left border of density of states which have a close to zero projection of magnetic moment $m$.
This situation corresponds to an antiferromagnetic behaviour.
For an accurate analysis of $g(E, m)$ the values of $\overline E$ and $\sigma^2$ depending on $m$ were calculated.

The formulas for the mean exchange energy (\ref{mean_E}) and variance (\ref{variance_E}) are written for $N \gg 1$.
Numerically, it is only feasible to model a system with a small number of spins $N$.
In this case, terms of the next order of magnitude in $N$ cannot be neglected.
In addition, each spin interacts only with the remaining $N-1$ spins. 
Taking these factors into account, the mean value of the total exchange interaction energy should be rewritten as follows:
\begin{equation} 
\begin{aligned} 
\overline {E} = \frac{1}{8} N \overline{J_1} \left( m^2 - \frac{1}{N} + O\left( \frac{1}{N^2} \right) \right)
\label{mean_EN}
\end{aligned}
\end{equation}

\begin{equation} 
\sigma^2 = (N-1) \sigma_1^2 \left(1 - \frac{2}{3} m^2 - \frac{1}{3} m^4 \right)
\label{variance_EN}
\end{equation}

\begin{figure}[!htbp]
\centering
\includegraphics[width=0.7\linewidth]{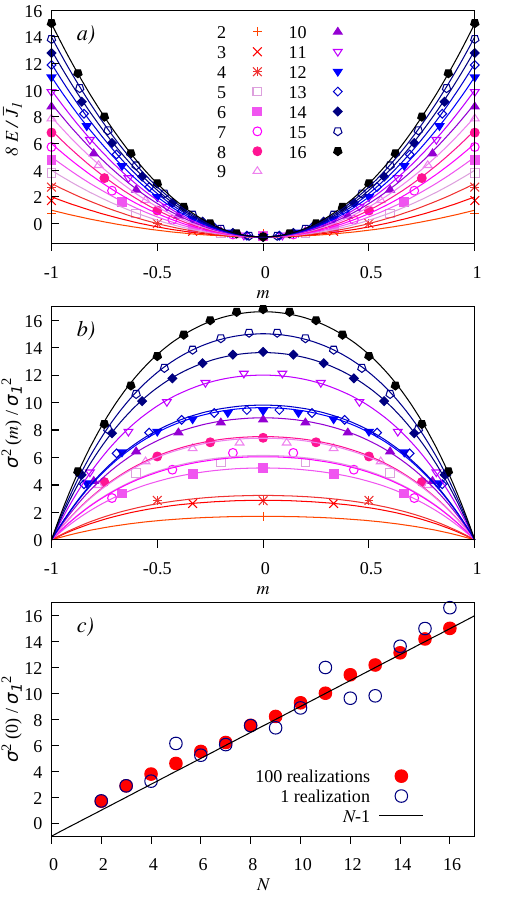}
\caption{$a)$ Dependence of the average energy $\overline{E}(m)$ for $N = 2\,$--16.
Dots show the computation results, lines --- analytical calculations by (\ref{mean_EN}).
$b)$ Dependence of the variance $\sigma^2(m)$ normalized to  $\sigma_1^2$ given by equation~\ref{variance_1}.
Dots show the computation results, lines are calculated analytically using equation (\ref{variance_EN}) where $\sigma^2(0)$ is a fitting parameter.
Figure $c)$ shows the dependence of the fitting parameter $\sigma^2(0)$ normalized to its theoretical value given by equation~\ref{variance_1} on $N$.
$\delta J=1\%, na^3=0.01$. When averaging over 100 realizations, the value $\delta J=5 \%$ was used.}
\label{moments}
\end{figure}

As can be seen at figure~\ref{moments}, the calculated $\overline{E}(m)$ is in close agreement with the formula (\ref{mean_EN}). 
The obtained $\sigma^2(m)$ are in reasonable agreement with the formula (\ref{variance_EN}), 
although the parameter $\sigma^2(0)$ fluctuates around the value predicted by the formula
(\ref{variance_1}). Nevertheless, after averaging $\sigma^2(0)$ over 100 realizations, good agreement
with the analytical formula was obtained.

It has to be noted that despite the agreement between analytical equations and numerical results for mean energy and variance,
higher moments as skewness and kurtosis are not nesessary equal to zero. It means that energy distribution $g(E, M)$ with
certain $M$ could deviate from normal distribution, especially on the tailes~\cite{PhysRevLett.89.070201}.
Therefore the Gaussian approximation, which takes into account only mean energy and variance, is apllicable only when temperature and size of the system are high enough.
Another reason for the inaccuracy of the Gaussian approximation at low temperatures is the discrete structure and finiteness of the density of states.

\section{Magnetic susceptibility}

We have calculated the mean value of the exchange energy and the variance. Now we can write an expression for the density of states of the system $g_q(E)$ at a given magnetic moment $M = \frac{1}{2} g \mu (N-2q)$. A macrostate with a total magnetic moment $M$ can be realized by $\binom{N}{q}$ number of microstates.
\begin{equation}
g_q(E) = \binom{N}{q}  \frac {1} {\sqrt {2 \pi } \sigma} 
\exp \left ( {- \frac { \left ( E- \overline {E} \right ) ^2} {2 \sigma^2}} \right ) 
\label{g1}
\end{equation}

For the antiferromagnetic interaction, the states for which $q \approx N/2$ are of greatest interest. In this case, the following expansion is valid

\begin{equation} \label{binom}
\binom{N}{q} \approx 2^N \sqrt{\frac{2}{\pi N}} \exp \left( - \frac{ \left( N-2q \right) ^2} {2N} \right)
\end{equation}

We substitute (\ref{binom}) into (\ref{g1})
\begin{equation}
g_q(E) = \frac {2^N } {\sqrt {N } \pi \sigma} 
\exp \left( - \frac{ \left( N-2q \right) ^2} {2N} - \frac { \left ( E- \overline {E} \right ) ^2} {2 \sigma^2} \right)  
\end{equation}

For further calculations, it will be convenient to non-dimensionalize the energy and variance.
We introduce a dimensionless energy per one spin $e = \frac{E}{N \overline{J_1}}$, and a dimensionless standard deviation $s_1 = \frac{\sigma_1}{\overline{J_1}}$. Since we are considering the antiferromagnetic case $m \ll 1$, we will omit terms of order $m^4$ in the expression for the variance.
\begin{equation}
\begin{aligned}
g(e,m) =  \frac {2^N} {\pi s_1 \sqrt{1 - \frac{2}{3}m^2  }} \times \\
\exp \left( - \frac{ N m^2 } {2} - \frac {N \left ( e - \frac{1}{8} m^2 \right ) ^2} {2 s_1^2 \left( 1 - \frac{2}{3}m^2 \right) } \right)  
\end{aligned}
\end{equation}

Now let us consider the system at temperature $T$ and in magnetic field $B$.
The probability of the system being in some state will be described by the Gibbs distribution 
with energy $E - MB = N \overline{J_1} e - \frac{1}{2} g \mu N m B$.
The probability density is
\begin{equation}
f(e,m,T,B) = \frac{1}{Z(T,B)} g(e,m) \exp \left( - \frac {N \overline{J_1} e  - \frac{1}{2} g \mu N m B}{kT} \right)
\label{Gibbs}
\end{equation}
Here $Z (T,B)$ is the partition function. For convenience, we introduce the dimensionless temperature $t = kT / \overline {J_1}$ and the dimensionless magnetic field $\beta = g \mu B / 2 \overline {J_1}$ and write out an explicit expression for the partition function.
%\begin{equation}
%\nonumber
%Z(t, \beta) =  \iint  g(e,m) \exp \left ( - \frac { Ne - N \beta m} {t} \right ) de dm
%\end{equation}
\begin{equation}
\nonumber
Z(t, \beta) =  \int\limits_{-1}^{+1} \int\limits_{-\infty}^{+\infty}  g(e,m) \exp \left ( - \frac { Ne - N \beta m} {t} \right ) de dm
\end{equation}

First, we integrate over energy
\begin{equation}
\begin{aligned}
\nonumber
Z(t, \beta) =  \int\limits_{-1}^{+1}  2^N \sqrt{\frac {2} {\pi N}} \times \qquad \qquad \\
\exp \left( - \frac{ N m^2 } {2 t^2}  \left( t^2 + \frac{t}{4} + \frac{2s_1^2}{3} \right) + \frac { N \beta m} {t} + \frac{ N s_1^2} {2 t^2} \right ) dm
\end{aligned}
\end{equation}

In the case of $N \gg 1$ and for weak magnetic fields $\beta \ll s_1^2/t$ the integral over $m$ from $-1$ to 1 can be replaced by the integral from $- \infty$ to $\infty$, 
which could be calculated analytically.
\begin{equation}
\nonumber
Z(t, \beta) = \frac {2^{N+1} t} { N \sqrt{ t^2 + \frac{t}{4} + \frac{2s_1^2}{3}}}
\exp \left( \frac{ N \beta^2 } {2  (t^2 + \frac{t}{4} + \frac{2s_1^2}{3})} + \frac{N s_1^2} {2 t^2} \right )
\end{equation}

The analytical expression for the probability density can be used to calculate various characteristics of a system of randomly distributed spins.
For example, the average magnetic moment of the system can be calculated as follows: 
$\overline {M}(t,\beta) = \frac{1}{Z(t,\beta)} \iint{M f(e,m,t,\beta)} de dm$.
For brevity, we introduce the notation $P (t,\beta) = \iint{m f(e,m,t,\beta)} de dm$ then $\overline {M} = \frac{g \mu N}{2} \frac{P}{Z}$.
The magnetic susceptibility of the system is
\begin{equation} 
\chi = \frac{1}{V}\frac{\partial \overline{M}}{\partial B} = 
\frac{g^2 \mu^2 n}{4 \overline{J_1}} \frac{\partial \overline{m}}{\partial \beta} = 
\frac{g^2 \mu^2 n}{4 \overline{J_1}} \left( \frac{1}{Z} \frac{\partial P}{\partial \beta} - \frac{P}{Z^2} \frac{\partial Z}{\partial \beta} \right)
\label{susceptibility}
\end{equation}

Here $V$ is the volume and $\overline{M}/V$ is magnetization of the spin system.
For large systems $N \gg 1$ the magnetic susceptibility in zero magnetic field is

\begin{equation} 
\chi  = \frac{g^2 \mu^2 n}{4 \overline{J_1} (t +\frac{1}{4} + \frac{2s_1^2}{3t})} 
\end{equation}

This expression can be compared with the results of low-temperature magnetic susceptibility measurements. For this purpose, it will be convenient to plot the dependence of the inverse susceptibility on temperature.

\begin{equation} 
\frac{1}{\chi} = \frac{4 k T + \overline{J_1}}{g^2 \mu^2 n} + \frac{8 \sigma_1^2}{3 g^2 \mu^2 n kT}
\nonumber
\end{equation}

In the hydrogen-like model $\sigma_1^2/\overline{J_1}^2 \ll 1$. Therefore, for the considered temperature range $kT > \overline{J_1}$ the last term can be neglected.
\begin{equation} 
\frac{1}{\chi} = \frac{4 k (T + T_N)}{g^2 \mu^2 n}
\label{curie}
\end{equation}
Here $T_N = \overline{J_1}/4 k$ is the Néel temperature. 
When comparing with experimental results, one can use the value obtained for the hydrogen-like model, or consider $T_N$ as a fitting parameter.

\section{Density of states approach for the Heisenberg spin chain}

In order to demonstrate the reliability and accuracy of the density of states approach and to
compare it with other numerical and analytical methods, we applied this method to calculate the
susceptibility of the Heisenberg spin ring, which is a closed spin chain, with antiferromagnetic interaction of spins. The spin
chains and spin rings have been thoroughly studied in the literature, both analytically and by
various numerical methods \cite{Bethe, PhysRev.135.A640, PhysRevB.22.5355}. The exact solution for
the magnetic susceptibility of the spin ring with the Ising interaction was obtained in
\cite{PhysRev.135.A640}.

The Hamiltonian of the Heisenberg spin ring has the form:
\begin{equation}
\vu{H} = \sum_{i} J_0 \vu{S}_i \vu{S}_{i+1} - g \mu B \sum_i \vu{s}_{zi}
\end{equation}

For rings of $N \le$ 16 spins we can exactly diagonalise the Hamiltonian and find all the states and
the corresponding energies. Thus, we can find the exact density of states and calculate the magnetic
susceptibility.
Furthemore, the mean value and variance of the total exchange energy can be calculated analytically similar to those described in Chapter II.

The mean value of the total exchange energy is
\begin{equation}
\begin{aligned}
\overline {E}_{ring} = N J_0 \overline {\vu{S}_i \vu{S}_{i+1}} = \frac{1}{4} N J_0 \left( m^2 + \frac{m^2}{N} - \frac{1}{N} + O\left( \frac{1}{N^2} \right) \right)
\label{avg_ring}
\end{aligned}
\end{equation}

And the variance is
\begin{equation}
\sigma^2_{ring} = \frac{1}{16} N J_0^2 \left(3 - 4m^2 + m^4 + \frac{2 - 6 m^2 + 4m^4}{N} + O\left( \frac{1}{N^2}\right) \right)
\label{variance_ring}
\end{equation}

\begin{figure}[!htbp]
\centering
\includegraphics[width=0.6\linewidth]{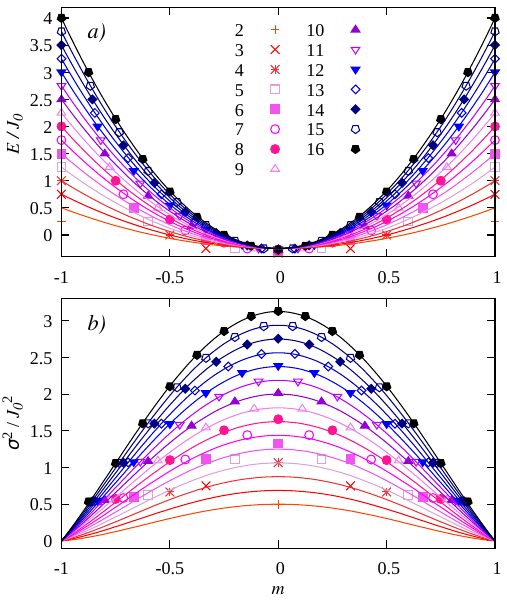}
\caption{The mean value and the variance of the exchange energy for a spin ring versus the magnetic moment $m$ calculated by the exact diagonalization (dots) and analytical formulas (\ref{avg_ring}, \ref{variance_ring}) (solid lines). Different colours correspond to different number os spins $N$ in the ring.}
\label{ring}
\end{figure}

The dependence of the mean value and the variance of the exchange energy on the magnetic moment is
shown in figure \ref{ring}. The dots are calculated by the exact diagonalization of the Hamiltonian
and the sollid lines are calculated by analytical equations (\ref{avg_ring}, \ref{variance_ring}).
As can be seen from the figures, the analytical equations match the results of the exact
diagonalization perfectly, even for small $N$.

\begin{figure}[!htbp]
\centering
\includegraphics[width=1.0\linewidth]{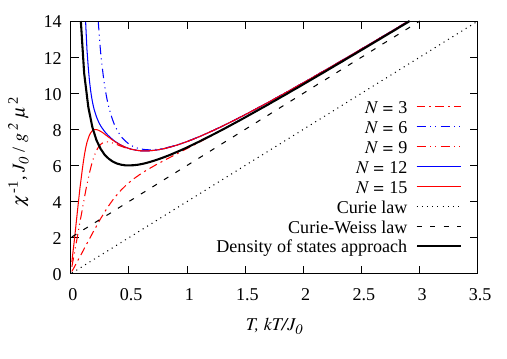}
\caption{Inverse magnetic susceptibility in zero magnetic field for rings of $N$ = 2..16 spins as a function of temperature.}
\label{chi2}
\end{figure}

For rings with $N \le$ 16 spins we calculated the magnetic susceptibility from the density of states obtained by exact diagonalization of the Hamiltonian. Similarly to \ref{Gibbs} we multiply the density of states by the Gibbs exponent and calculate the magnetic moment in as a function of the magnetic field.
The dependence of the inverse susceptibility on temperature is shown in the figure~\ref{chi2} for various $N$. 
At relatively high temperatures $t > 1$ the dependences for all $N$ almost coinside. 
However, at low temperatures the results obtained for even and odd $N$ are significantly different.
For even $N$, the spins are split into pairs and the magnetic susceptibility is small. 
For odd $N$ the total magnetic moment of the system is always non-zero, so the magnetic susceptibility is large. 

The magnetic susceptibility for a ring of $N \gg 1$ spins was calculated analytically by the method described in Chapter IV. For this, we discard the terms of the order $1/N$ in (\ref{avg_ring}, \ref{variance_ring}) and obtain the following expression for the inverse susceptibility per one spin in zero magnetic field:

\begin{equation}
\frac{1}{\chi} = \frac{4 k T + 2 J_0}{g^2 \mu^2 } + \frac{J_0^2}{g^2 \mu^2 kT}
\label{chi_ring}
\end{equation}

At low temperatures the susceptibility decreases as for an even-numbered spin ring.
In the figure \ref{chi2}, this dependence is shown by a red solid line. It can be seen that at temperatures $t > 2$ this line agrees perfectly with the results obtained by the exact diagonalization method. It is significant that both methods significantly differ from the simple Curie law for the susceptibility of free spins ($\frac{1}{\chi} = \frac{4 k T}{g^2 \mu^2 }$, black line) and the Curie-Weiss law ($\frac{1}{\chi} = \frac{4 k T + 2 J_0}{g^2 \mu^2 }$, purple line).
In the last case we omit the last term in (\ref{chi_ring}).
Good agreement with the results of the exact diagonalization at relatively high temperatures confirms the applicability of the density of states approach.

Now let us compare the obtained results with the results known from the literature. 
The magnetic susceptibility for a spin ring with an antiferromagnetic interaction was analytically and numerically calculeted in \cite{PhysRev.135.A640}. Our results obtained by the exact diagonalization method are in excellent agreement with the results obtained in \cite{PhysRev.135.A640}. 

As could be expected, the density of states method based on statistical averaging agrees with the exact diagonalization method at relatively high temperatures $t > 1$. 
At lower temperatures the maximum of the probability density correspond to the edge of the density of states, 
where the distribution differs from the normal one. 
At even lower temperatures, the maximum of the probability density for a normally distributed density of states shifts to lower energies where there are no states. Therefore, the density of states method overestimates the magnetic moment of the system and the magnetic susceptibility at low temperatures. This explains why at temperatures $t<1$, the susceptibility calculated by (\ref{chi_ring}) is higher than obtained by the exact calculation (see Fig. \ref{chi2}).
At intermediate temperatures $t \approx 1$ the strong disorder renormalization group method \cite{PhysRevLett.48.344} is applicable.

\section{Comparison with the experimental data}
\begin{figure}[!htbp]
\centering
\includegraphics[width=1\linewidth]{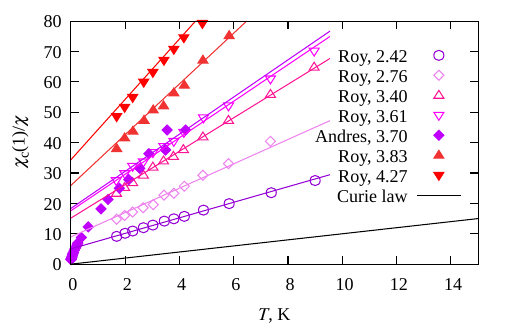}
\includegraphics[width=1\linewidth]{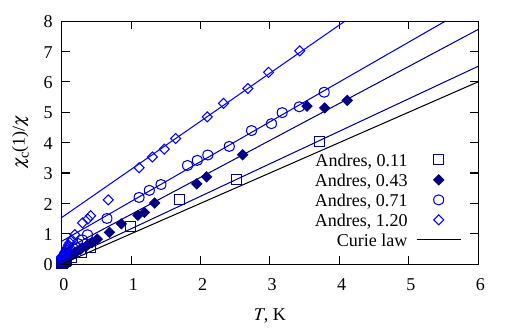}
\caption{Comparison of experimental data from Roy et al.~\cite{PhysRevB.37.5531} and Andres et al.~\cite{PhysRevB.24.244} with the fitted results. Concentration is given in $10^{18}\,$cm$^{-3}$ }
\label{chi_exp}
\end{figure}

In the works of Roy et al.~\cite{PhysRevB.37.5531} and Anders et al.~\cite{PhysRevB.24.244}, measurements of the magnetic susceptibility of phosphorous impurities in silicon at low temperatures were conducted.
The figure~\ref{chi_exp} shows the measured inverse susceptibility $\chi_C(1)/\chi$, normalized to the Curie susceptibility at $1\,$K which is numerically equal to $g^2 \mu^2 n / 4 k$. Solid lines represent the results of fitting using the following formula:
\begin{equation} 
\frac{\chi_C(1)}{\chi(T)} = \frac{T + T_{exp}n_0}{C_i},
\label{chiexp}
\end{equation} 
where $T_{exp}$ is regarded as a fitting parameter, the same for all experimental dependences. 
In comparison to the theoretical formula (\ref{curie}), $T_{exp}$ corresponds to $T_N(n = 10^{18}\,$cm$^{-3})$,
$n_0$ is a dimensionless concentration which is expressed in units of $10^{18}\,$cm$^{-3}$,
and $C_i$ is a fitting parameter, which represents the fraction of electrons localized at impurity centers and varies from zero to one. 
Indeed, it is known that amount of localized magnetic moments is changing while the impurity concentration is close to
the metal-insulator transition. Both theoretically~\cite{PhysRevLett.63.82} and experimentally~\cite{PhysRevB.37.5531}, it was demostrated that the localized magnetic moments
exist even on the metallic side of transition. Review of the problem can be found in~\cite{RevModPhys.66.261}.

In accordance with the formula (\ref{curie}), the dependence $1/\chi(T)$ should be parallel to $1/T$.
However, in the experiment, the slope also changes as the concentration increases.
This phenomenon can be explained by the fact that in the vicinity of the metal-insulator transition concentration, a considerable part of electrons undergoes a transition to the metallic phase.
The susceptibility of electrons in the metallic phase is independent of temperature and is significantly lower than that of localized electrons. Consequently, the contribution of delocalized electrons can be neglected. This behavior can be taken into account by the introduction of the fitting coefficient $C_i$.

\begin{figure}[!htbp]
\centering
\includegraphics[width=1\linewidth]{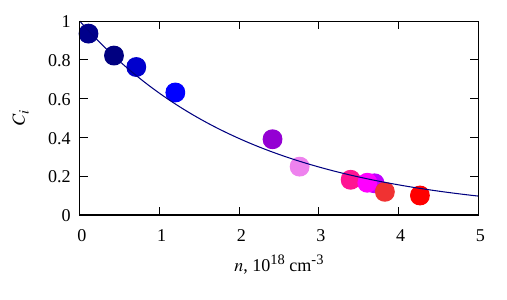}
\caption{Dependence of the fraction of localized electrons $C_i$ on the concentration of impurity centers. The solid line is an empirical fit using the formula $C_i = \exp(-n/n_c), n_c\approx2.15\cdot10^{18}\,$cm$^{-3}$. The color of the dots correspond to the color in figure~\ref{chi_exp} with the same impurity concentraion. }
\label{C_n}
\end{figure}

The fitting procedure was conducted as follows.
All experimental points for which the condition $T> 0.7\,T_{exp} n_0$ was satisfied were fitted by linear dependencies.
The coefficient $0.7$ was chosen empirically to cut off the low-temperature range, within which the transition to the Bhatt-Lee random singlet phase~\cite{PhysRevLett.48.344} occurs. At this temperature range, the strong disorder renormalization group approach is applicable.
The parameter $T_{exp}$ was varied in order to minimize the sum of squares of deviations from the experimental data for all curves.
Figure \ref{C_n} shows the obtained $C_i(n)$ dependence. 
It is evident that it has the expected form, i.e. at low concentrations it tends to 1,
whereas at concentrations close to the concentration of the metal-insulator transition it goes to zero.

In the hydrogen-like model~\cite{1964Gorkov, PhysRev.134.A362} the parameter $J_0$ in (\ref{J_r}) is $1.641 Ry$.
In the paper~\cite{PhysRevB.24.244} the following values are given for P impurity in Si: $Ry \approx 45.5\;$meV and the Bohr radius $a_B \approx 1.52\;$nm.
Consequenly $T_N =\overline{J_1} / 4k = 10.9\;$K.
Upon fitting the experimental data, a significantly lower value was obtained $T_{exp} = 0.8076\;$K.
It should be noted, however, that the conduction band minimum in Si consists of six valleys, and only states from one valley interact.
Therefore, the theoretical value should be reduced by a factor of six $T_N/6 = 1.82\;$K. 
This value remains approximately twofold that obtained from the fitting. 
The exchange interaction between impurities in silicon has been the subject of extensive investigation in the literature. For instance, the hydrogen-like model has been shown to significantly overestimate the exchange interaction between two impurity atoms in Si~\cite{PhysRevB.91.235318, PhysRevB.68.155206}.
Our results are consistent with this conclusion.

Although the hydrogen-like model significantly overestimates the strength of the exchange interaction,
good agreement between the theoretical curves and the experimental results indicates that the proposed model 
provides good qualitative agreement with the experiment, and for quantitative agreement, 
the value of $T_N$ should be considered as a fitting parameter.

\section{Magnetization in strong magnetic fields}

In our previous paper~\cite{PhysRevB.109.024436}, the density of states approach was applied to a disordered system of spins with a ferromagnetic exchange interaction.
In this case, the magnetic moment $m$ can be close to~1, and thus the expansion (\ref{binom}) is not applicable.
Instead, the binomial coefficient was expanded using the Stirling formula.
For the antiferromagnetic exchange interaction in a strong magnetic field, the average magnetic moment $m$ can also be quite large.
Using the Stirling formula, the binomial coefficient can be written as
\begin{equation}
\nonumber
\binom{N}{\frac{N(1-m)}{2}} = \sqrt{\frac{2}{\pi N}} \frac {1}{\sqrt{1 - m^2}} \exp \left( N p(m) \right)
\end{equation}
Here, for brevity, we use the notation 
\begin{equation}
\nonumber
p(m) = \ln 2 - \frac{1-m}{2} \ln(1-m) - \frac{1+m}{2} \ln(1+m)
\end{equation}

Then, using the approach proposed in ~\cite{PhysRevB.109.024436}, we obtain the following expression for the partition function:
\begin{equation}
\begin{aligned}
Z(t, \beta) = \sqrt{\frac{N}{2 \pi}}  \int \frac{dm}{\sqrt{1 - m^2}}  \times \qquad \qquad \\
\exp \left ( N \left( p(m) + \frac {s_1^2 (3 - 2m^2 - m^4)}{6 t^2} - \frac{m^2}{8t} + \frac{\beta m}{t} \right)\right)
\label{partition}
\end{aligned}
\end{equation}

The integral over the magnetic moment can be calculated using the Laplace's method.
The maximum value of the exponent in equation (\ref{partition}) is achieved at $m = m_0$, which is determined from the following transcendental equation:

\begin{equation} \label{m0}
- \frac{2 s_1^2 m_0}{3 t^2} - \frac{2 s_1^2 m_0^3}{3 t^2} - \frac{m_0}{4 t} + \frac{\beta}{t} + \frac{1}{2} \ln \left( \frac{1-m_0}{1+m_0} \right) = 0
\end{equation}

In the case of large systems $N \gg 1$, the maximum of the exponent is markedly sharp, therefore states with a magnetic moment that differs from $m_0$ are essentially unattainable.
Thus, equation (\ref{m0}) provides the magnetic moment of the system.
At relatively high temperatures and relatively weak magnetic fields, the last two terms in equation (\ref{m0}) are the main ones, and the logarithm can be expanded in a series in the small parameter $m_0$. This allows to reduce the equation (\ref{m0}) to the form $m_0~=~\beta/t$. When $m_0$ is not small, equation (\ref{m0}) can be solved numerically.

In~\cite{PhysRevB.34.387}, experimental measurements of impurity magnetization $\mathcal{M}$ for Si samples doped with different concentration of P are presented.
The so-called scaling behavior of magnetization was found.
Specifically, the magnetization in the scaling coordinates ($\mathcal{M}(B,T)/T \chi(0, T)$; $B/T$) was found to be independent of the sample temperature.
This phenomenon was theoretically explained within the framework of the Bhatt and Lee model~\cite{PhysRevLett.48.344} and was associated with the power-law dependence $\chi(T)\sim T^\alpha$.
Our model also predicts similar scaling behavior in the temperature range $T>T_N$.

Indeed, at high temperatures the magnetzation of the system $\mathcal{M}(B,T) = \frac{1}{2} n g \mu m_0 \approx n g \mu \beta/ 2 t = n g \mu B/2kT$.
Magnetic susceptibility is $\chi(T) \approx g^2 \mu^2 /4kT $.
Therefore, in the high temperature limit $\mathcal{M}(B,T)/T \chi(0, T) \approx 2B/gT \approx B/T$ taking into account that the spin g-factor is equal to 2.
This explains the scaling behaviour experimentally observed in~\cite{PhysRevB.34.387} at low magnetic fields and high temperatures.
In the case of strong magnetic fields, the magnetization and magnetic susceptibility can be determined by numerically solving equation (\ref{m0}).

\begin{figure}[!htbp]
\centering
\includegraphics[width=1\linewidth]{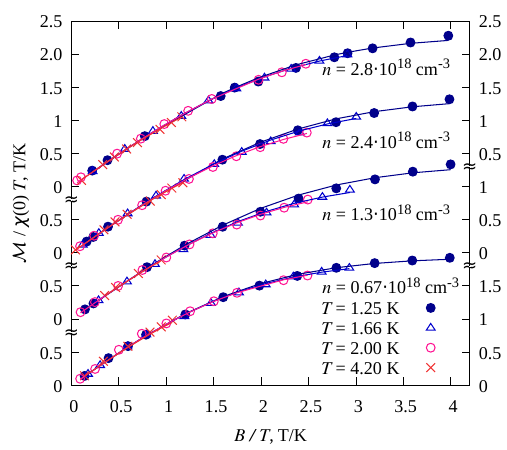}
\caption{Dependences of magnetization on magnetic field in scaling coordinates. 
Dots are experimental data from~\cite{PhysRevB.34.387}; solid lines calculated by the formula~(\ref{С_fs})}
\label{scaling}
\end{figure}

\begin{figure}[!htbp]
\centering
\includegraphics[width=1\linewidth]{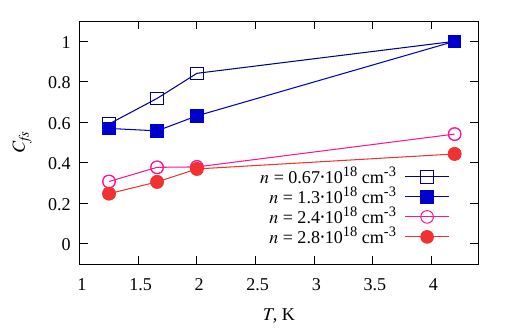}
\caption{Dependences of the fraction of free spins $C_{fs}$ on temperature for samples with different impuripy concentrations, calculated based on fitting the experimental curves from~\cite{PhysRevB.34.387}}
\label{Cfs}
\end{figure}

This equation is written for a hydrogen-like exchange interaction.
However, such a model significantly overestimates the magnitude of the exchange interaction.
The average exchange energy (\ref{j1}) and variance (\ref{variance_E}) depend on the parameter $J_0$.
Based on the results of fitting the inverse susceptibility on temperature, 
the value of $J_0$ should be reduced by a factor of $T_{exp}/T_{N18}$ in comparison to the hydrogen-like model. 
Here $T_{N18}$ is the Néel temperature at concentration $n=10^{18}\,$cm$^{-3}$. 
Taking this into account, equation (\ref{m0}) can be rewritten as follows:

\begin{equation}
\begin{aligned}
\label{С_fs}
C_{fs} n_0 \left(- \frac{2 s_{18}^2 m_0}{3t^2} - \frac{2 s_{18}^2 m_0^3}{3t^2} - \frac{m_0}{4t}\frac{T_{exp}}{T_{N18}}\right) + \\
+\frac{\beta}{t} + \frac{1}{2} \ln \left( \frac{1-m_0}{1+m_0} \right) = 0  \qquad \qquad
\end{aligned}
\end{equation}
Here $s_{18}$ denotes the value of the dimensionless standard deviation $s_1$ at a concentration $n = 10^{18}\,$cm$^{-3}$, and the dependence on concentration, the same for the first three terms, is taken out of brackets.
The parameter $C_{fs}$ denotes the fraction of free spins that have not coupled into singlets.
Indeed, the magnetization measurements in the paper~\cite{PhysRevB.34.387} were carried out at low temperatures comparable to the Néel temperatures for the studied dopant concentrations.
Therefore, a part of the spins will couple into singlet pairs with zero magnetic moment, which will manifest itself as an effective decrease in concentration.

The dots in figure~\ref{scaling} show experimental dependencies from~\cite{PhysRevB.34.387} in scaling coordinates.
The solid curves represent the result of fitting using formula (\ref{С_fs}).
Good quantitative agreement between the calculation and experimental data is evident.

Figure~\ref{Cfs} shows the dependence of the fitting coefficient $C_{fs}$ on temperature for samples with different concentrations.
As expected, at low concentrations and high temperatures $C_{fs} \approx 1$.
As the concentration increases and the temperature decreases, the fraction of free spins decreases. 
This dependence is qualitatively consistent with the Bhatt-Lee concept of singlet spin pairs formation.
This shows that the density of states approach is applicable not only for temperatures $T>T_N$, but also can be used for lower temperatures;
at low tempetatures this approach allows for the estimation of the fraction of spins coupled into singlets.

\section{Conclusion}
This paper presents a study of the magnetic properties of doped semiconductors using the density of states approach.
The magnetic susceptibility and magnetic moment were calculated in a wide range of magnetic fields and temperatures.
The obtained dependences are used to describe the experimental results published in the literature.
From the fitting of the experimental data, the fraction of localized magnetic moments as a function of concentration, as well as the ratio of free electrons and electrons coupled into singlets, depending on concentration and temperature, were estimated.
The quantitative correspondence between the theory and the experiment, as well as the qualitative agreement with the results of calculations performed within the framework of the strong-disorder renormalization group, are demonstrated.
The results obtained are of importance for the development of the theory of disordered antiferromagnets and for a deeper understanding of the physics of the metal-insulator transition in doped semiconductors.

\section{Acknowledgements}
We acknowledge support from Russian Science Foundation (Grant No. 23-22-00333).

\newpage

%\bibliography{article}% Produces the bibliography via BibTeX.


\end{document}